\documentclass[12pt]{article}
\usepackage{epsfig,url}
\newcommand{\ba}{\begin{eqnarray}}
\newcommand{\ea}{\end{eqnarray}}
\renewcommand{\kappa}{{k}}
\def\lsim{\mathrel{\raise.3ex\hbox{$<$\kern-.75em\lower1ex\hbox{$\sim$}}}}
\def\gsim{\mathrel{\raise.3ex\hbox{$>$\kern-.75em\lower1ex\hbox{$\sim$}}}}
\begin{document}
\begin{titlepage}

\begin{centering}
\begin{flushright}
hep-th/0205299
\end{flushright}

\vspace{0.1in}

{\Large {\bf Exact Solutions and the Cosmological Constant Problem in
Dilatonic-Domain-Wall Higher-Curvature String Gravity}}
\vspace{0.1in}

{\bf Nick E. Mavromatos } \\
Department of Physics, Theoretical Physics, King's College London,\\
Strand, London WC2R 2LS, United Kingdom.

\vspace{0.05in}
and
\vspace{0.05in}

{\bf John Rizos} \\
Department of Physics, University of Ioannina, \\
GR 45100 Ioannina, Greece,\\

\vspace{0.1in}
 {\bf Abstract}

\end{centering}

\vspace{0.05in}

{\small In this article we extend previous work by the authors,
and elaborate further
on the structure of
the general solution to the graviton and dilaton equations
of motion in brane world scenaria,
in the context of
five-dimensional
effective actions with ${\cal O}(\alpha ')$
higher-curvature corrections,
compatible with bulk string-amplitude calculations.
We consider (multi)brane scenaria, dividing the bulk space into regions,
in which one matches two classes of general solutions,
a linear Randall-Sundrum solution and a (logarithmic) dilatonic domain wall
(bulk naked singularity).
We pay particular attention to examining
the possibility of resolving the mass hierarchy problem together
with the vanishing of the vacuum energy on the observable world,
which is taken to be a
positive tension brane.
The appearance of naked dilatonic domain walls provides
a dynamical restriction of the bulk space time.
Of particular interest is a dilatonic-wall solution, which after
appropriate coordinate transformation, results in a linear
dilaton conformal field theory. The latter may provide
a holographic resolution of the naked singularity problem.
All the string-inspired models involved
have the generic feature that the brane tensions are
proportional to the string coupling $g_s$; it remains a
challenge for string theory, therefore, to show whether
microscopic models respecting this feature can be constructed.}

\vspace{0.4in}
\begin{flushleft}
May 2002
\end{flushleft}

\end{titlepage}

\newpage
\section{Introduction}

Randall and Sundrum~\cite{RS} (RS) have proposed
a new solution to the mass hierarchy problem
on the observable world
by means of gravitational interactions
in the bulk coordinate of a five-dimensional
space time, in which our world is viewed as an embedded
three-brane domain wall. The scenario
involves a hidden world, represented
by a second parallel brane, whose separation
from the first determines the mass hierarchy
on the observable world.  A necessary ingredient in the
above construction is an orbifold in the bulk direction,
which restricts the bulk space onto that between the branes.
This set-up is implemented by the following
non-factorizable form of the metric:
\ba
 ds^2 = e^{-2\sigma(z)} \eta_{ij}dX^i dX^j + dz^2~, i,j=0,1,\dots 3
\label{RSmetric}
\ea
This is a solution to the standard Einstein equations
provided one chooses the
following
form of the metric function $\sigma(z)$,
in the simplest case of a constant dilaton $\Phi$:
$\sigma (z) = \sum _{i} \kappa \left|z - z_i\right|$,
with $i$ denoting the $i-th$ brane, located at $z_i$ along the
bulk direction. This metric is known in the literature
as the
Randall-Sundrum (RS) metric~\cite{RS}.

In a recent article~\cite{mr},
we have considered
five-dimensional actions including
${\cal O}(\alpha ')$ higher-curvature corrections,
compatible with string-amplitude computations~\cite{string},
with non-constant (bulk-coordinate dependent) dilaton fields, where
$\alpha' =M_s^{-2}$ is the Regge slope of the string,
and $M_s$ the
string mass scale.
Such corrections, but for constant
dilaton fields or for actions that are not
derived from string amplitudes, had been
considered earlier in \cite{zee,otherszee}, with
different conclusions from ours.

The appearance of quadratic (Gauss-Bonnet type) terms in the
effective action, may occur in Heterotic M-theory scenaria of
Horava-Witten type~\cite{heterotic}, which are known to have
D-branes. In type IIB string theories on the other hand, which are
also known to to admit D-brane solutions, there are no quadratic
Gauss-Bonnet corrections at tree level, although such terms may be
generated through loop corrections. The presence of
higher-curvature terms in the effective action leads to
interesting physics, and a lot of works have appeared recently
examining potential physical applications in a variety of context,
ranging from static solutions to cosmology~\cite{allGBs,allGBc,allGBa}.
Moreover, in standard string theory, it is known that
higher-curvature corrections lead to highly non-trivial results,
for instance new types of Black-Hole solutions with (secondary)
dilaton hair~\cite{kanti}, or singularity-free
cosmologies~\cite{art}.

In view of this wide range of applications, we therefore feel that
it is appropriate to present a systematic analysis of the general
structure of the solutions of the string-effective action to
${\cal O}(\alpha ')$, which
extends and completes the analysis of \cite{mr}.
In this article we
elaborate further on the structure of the
general solution to the equations of motion, and discuss
physical applications, especially from the point of view
of resolving simultaneously the mass hierarchy problem, the
positivity of the tension of our observable brane world,
and the vanishing of the
total vacuum energy induced on our world.

In particular, we examine two classes of solutions, a
linear Randall-Sundrum solution, with constant dilaton,
and a logarithmic domain-wall solution, where the bulk dependent
dilaton and the metric have bulk naked singularities.
The article deals with a discussion on the integrability conditions
of such singularities, as well as a
detailed matching of these classes
of solutions in various scenaria, where brane junctions separate
the bulk space into regions. The presence of dilatonic domain walls
implies a dynamical restriction of the bulk space.

The structure of the article is as follows:
In section 2 we review the general solution of \cite{mr}
to the graviton and dilaton equations of motion,
including the string-inspired
Gauss-Bonnet quadratic curvature contributions
and dilaton four-derivative terms in the gravitational action.
In section 3 we discuss effective quantities, such as
Planck mass and Vacuum energy,
as measured by a
four-dimensional observer, living on the observable
brane world.
In section 4 we discuss the structure of two types of exact
solutions.
One type is the Linear RS solution~\cite{RS}, while the
other is a dilatonic domain wall (bulk logarithmic naked singularity).
We discuss integrability conditions of these latter naked
singularities in the bulk space, to which
we restrict ourselves throughout this work.
We show how such integrable
singularities are consistent solutions of the ${\cal O}(\alpha ')$
equations of motion.
In section 5 we discuss
matching of these two types of exact solutions
at various brane junctions, separating
the bulk space into regions.
We commence our analysis
by discussing singe brane scenaria and their
disadvantages from a physical point of view.
We then proceed to analyze  multibrane scenaria
which can solve the mass hierarchy problem on
a positive tension observable world simultaneously
with a vanishing vacuum energy (cosmological constant) on the brane.
In section 6 we discuss a possible resolution of the bulk naked
singularity problem by means of holography induced by
a linear-dilaton domain-wall solution,
which, as we show, is a specific case of the general solution of \cite{mr}.
Finally, in section 7
we present our conclusions and outlook.

\section{General Solution with Gauss-Bonnet Interactions: a Review}

In this section we review the general solution of the ${\cal O}(\alpha ')$
string-inspired effective action, in the five-dimensional effective case,
discussed in detail in \cite{mr}.
We consider the action:
\ba
S= S_5 + S_4
\label{s5s4}
\ea
where $S_5$:
\ba
 S_5 &=&
\int d^5x \sqrt{-g} \left[ -R -\frac{4}{3}\left(\nabla_\mu \Phi \right)^2
+ f(\Phi) \left(R^2 -4 R_{\mu\nu}^2 +
R_{\mu\nu\rho\sigma}^2\right) \right.\nonumber \\
&~& + \xi(z) e^{\zeta \Phi}
+ \left. c_2~f(\Phi)\left(\nabla _\mu \Phi \right)^4 + \dots \right]
\label{actionGB}
\ea
with $\Phi$ the dilaton field, and
the $\dots$ denoting other types of contraction
of the four-derivative dilaton terms which will not be of
interest to us here, given that by appropriate field redefinitions,
leaving the string amplitudes invariant~\cite{string},
one can always cast such terms in the above form.

The four-dimensional part $S_4$ of the action (\ref{s5s4})
is defined as:
\ba
 S_4 = \sum_{i} \int d^4x \sqrt{-g_{(4)}} e^{\omega \Phi} v(z_i)
\label{s4}
\ea
where
\ba
g_{(4)}^{\mu\nu}=\left\{
\begin{array}{l}
g^{\mu\nu}\ , \,\mu,\nu<5\\
0 \ \ \ ,\ \mbox{otherwise}\\
\end{array}\right.
\ea
and the sum over $i$ extends over D-brane walls located at $z=z_i$
along the fifth dimension.

The above action is compatible with closed  string amplitude computations
in the five-dimensional space times~\cite{string}. In our
opinion, such a compatibility is not only natural, but also necessary
in view of the assumption of closed
string propagation in the bulk.
In the stringy case one has~\cite{mr}:
\ba
&~&~~f(\Phi)=\lambda
~e^{\theta\Phi}~,~
\lambda =\alpha '/8g_s^2 > 0~,~c_2=\frac{16}{9}\frac{D-4}{D-2} \nonumber \\
&~&\zeta=-\theta=\frac{4}{\sqrt{3(D-2)}}~~(=4/3~~ {\rm for}~~ D=5)
\label{lambdastring}
\ea
where $g_s$ is the string coupling and $D$ is the number of space-time
dimensions. In this work we shall restrict ourselves to the case
$D=5$~\cite{mr}. However, for reasons that will be clarified below,
we shall keep the coefficient $c_2$
of the dilaton quartic terms in the action general.
As we shall see, our results do not depend crucially on its value.

In this case, the equations of motion for the graviton and dilaton
fields obtained from (\ref{s5s4}) read respectively:
\ba
&~& 0= R^{\mu\nu} + \frac{1}{2}g^{\mu\nu}
\left(-R - \frac{4}{3}(\nabla\Phi)^2+c_2 f(\Phi) (\nabla\Phi)^4
+ \xi(z) e^{\zeta \Phi}\right) \nonumber \\
&~&+\frac{1}{2}\sum_i{\frac{\sqrt{-g_{(4)}}}{\sqrt{-g}}}g_{(4)}^{\mu\nu}
e^{\omega \Phi} v(z_i)+ \frac{4}{3}(\nabla^\mu \Phi) (\nabla^\nu \Phi)
  \nonumber \\
&~&-f(\Phi) \left(2\alpha R R^{\mu\nu} + 2\beta {R^\mu}_\sigma R^{\nu\sigma}
+ 2\gamma R^\mu_{\sigma\tau\rho}R^{\nu\sigma\tau\rho} \right)
\nonumber \\
&~& +\frac{1}{2} g^{\mu\nu} f(\Phi) \left(\alpha R^2 +
\beta R_{\sigma\tau}R^{\sigma\tau} + \gamma
R_{\sigma\tau\rho\kappa}R^{\sigma\tau\rho\kappa} \right) \nonumber \\
&~& + 2\alpha \{ {(g^{\mu\nu} f(\Phi)R)_{;\sigma}}^\sigma -
(f(\Phi)R)_;^{\mu\nu} \}  \nonumber \\
&~& +\beta \{
(g^{\mu\nu}f(\Phi)R^{\sigma\tau})_{;\sigma\tau} +
{(f(\Phi) R^{\mu\nu})_{;\sigma}}^\sigma
- {{(f(\Phi) R^{\mu\sigma})_;}^\nu}_{\sigma}
- {{(f(\Phi) R^{\nu\sigma})_;}^\mu}_{\sigma}\}  \nonumber \\
&~& + 2\gamma \{ (f(\Phi) R^{\mu\sigma\nu\tau})_{;\sigma\tau} +
(f(\Phi) R^{\mu\sigma\nu\tau})_{;\tau\sigma} \} \nonumber\\
&~&-  2c_2 f(\Phi) (\nabla^\mu\Phi) (\nabla^\nu \Phi)
{(\nabla \Phi)}^2
\label{gravitoneq}
\ea
and
\ba
&~&0= \frac{8}{3}\nabla^2 \Phi + f'(\phi)
\left(\alpha R^2
+ \beta R_{\mu\nu}R^{\mu\nu} + \gamma R_{\mu\nu\rho\sigma}R^{\mu\nu\rho\sigma} \right)+ \zeta \xi(z) e^{\zeta\Phi}+\nonumber \\
&~&\sum_i{\frac{\sqrt{-g_{(4)}}}{\sqrt{-g}}}\omega e^{\omega \Phi} v(z_i)
 - 4 c_2  \nabla_\mu \left(f(\Phi) (\nabla^\mu \Phi) (\nabla \Phi)^2 \right)
+c_2 f'(\Phi) \left(\nabla \Phi\right)^4
\label{dilatoneq}
\ea
In the above formul\ae \, the symbol $;$ denotes covariant differentiation, and
the prime denotes differentiation with respect to $\Phi$.

\begin{figure}[!t]
\begin{center}
\epsfig{figure=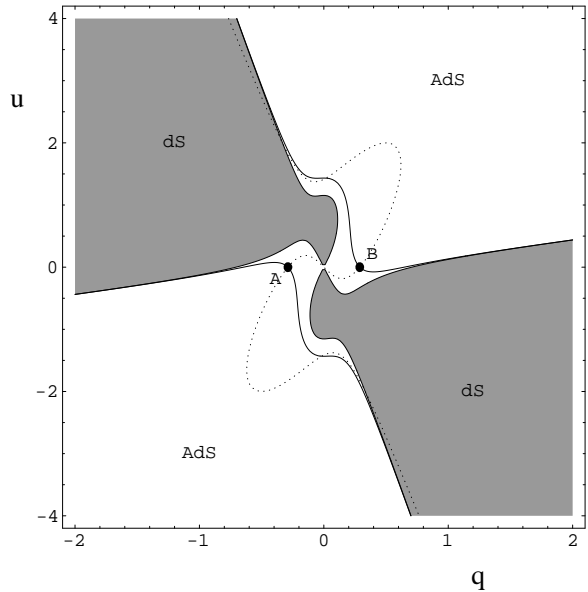, width=0.45\textwidth}
\caption{{\it The phase-space of the general solution
of the graviton-dilaton equations of motion
in the presence of
Gauss-Bonnet curvature-squared terms in the low-energy
effective action. The regions where the bulk cosmological
constant is negative (positive),
and hence the space time is de Sitter (dS) (anti-de-Sitter (AdS))
are indicated. For detailed explanations we refer the reader to
ref. \cite{mr}. The dashed lines indicate the exact logarithmic solution
(\ref{lowzeesol2}), while the dots at A and B denote the Randall-Sundrum (RS)
solution (\ref{RSexact}).
The continuous lines passing through A and B represent the
solution  of \cite{mr} interpolating between the RS solution
near our brane world
and a
bulk naked singularity at infinity. The integrability conditions
on the naked singularities impose further restrictions on the
solutions phase space to be discussed in the current article.}}
\label{fig:solutions}
\end{center}
\end{figure}

The general solution
has been studied in detail in \cite{mr}.
It has been shown there that among the solutions
there is one which interpolates {\it continuously}
between an RS type of solution (\ref{RSmetric}),
occurring at $z=+\infty$, and a naked singularity at $z=0$,
which is however integrable.
This solution is important in that it implies the
{\it dynamical} formation of domain walls
in the bulk direction, however
the relevant distance from the singularity
is infinite, and hence it cannot provide
a solution to the hierarchy problem {\`a} la
Randall-Sundrum~\cite{RS}.
Graphically, the phase space of the solution is
depicted in figure \ref{fig:solutions}.

The implementation of bulk naked singularities,
or equivalently dilatonic domain walls
as they are called in the literature~\cite{naked,kachru,naked2,nilles,youm},
due to the fact that
both dilaton and graviton fields exhibit logarithmic singularities,
has been connected
to a possible
resolution of
the cosmological constant problem
in the context of lowest-order
brane-world Einstein gravity~\cite{naked,kachru}.
The idea is that the vacuum energy of the brane world
curves the fifth dimension, leaving a flat
(Poincare invariant) four-dimensional
brane world intact.
However, as demonstrated in~\cite{nilles},
such a scenario hides a fine-tuning in the following sense:
one should resolve the naked singularity in a way
consistent with classical equations of motion.
This
requires the brane world tension  to be equal in magnitude
and opposite in sign with the tension of a second brane
(hidden world), placed at the singularity, whose presence
is necessitated by the requirement of satisfaction of
the equations of motion. In this way, the
expectations of~\cite{kachru} for a self tuning
mechanism for the cancellation of the cosmological
constant do not work.

As discussed recently in~\cite{davis},
if one imposes supergravity in the bulk,
then there are topological obstructions,
associated with extra degrees of freedom
of the supergravity multiplet, which force
the naked singularity to lie beyond the second brane,
and thus being shielded.

An additional problem with brane world scenaria is
the fact that in the original RS model~\cite{RS}
the physical world was located on a brane with negative
tension, which is a serious instability~\cite{rubakov}.
To tackle this problem, bigravity scenaria have been
proposed~\cite{bigravity}, i.e. multibrane scenaria
involving branes with alternating-sign tensions, in such a way that our
world is  identified with a positive tension brane surrounded by
negative tension ones, lying asymmetrically
from our world. Again such models have been
studied in the context of the lowest-order Einstein gravity actions,
and actually with constant dilaton fields in the bulk.
Moreover the original orbifold construction of \cite{RS} has been
imposed.

It is the purpose of this article to discuss
all the above issues, when string-inspired ${\cal O}(\alpha ')$
higher-curvature corrections are taken into account, with non-trivial
bulk dependence of the dilaton field, which, together with the graviton,
is assumed propagating in the bulk.
In particular,
we shall demonstrate first that the original Randall-Sundrum
scenario, with or without orbifold constructions,
is capable of solving both the hierarchy,
and the smallness of the cosmological constant on the brane world.
We shall also discuss other scenaria involving
dilatonic domain walls, specifically
we shall discuss
multibrane scenaria
involving  dilatonic brane worlds
shielded by a number of ghost branes, surrounding our physical brane world.
In such constructions
we shall demonstrate the possibility of solving
the mass hierarchy problem, on a positive tension observable brane world,
simultaneously with the vanishing of the vacuum energy
on the brane.
To this end, we shall employ alternative solutions
to the system of equations of motions (\ref{gravitoneq}),(\ref{dilatoneq}),
that were also presented in \cite{mr}, but not studied further there.
Finally, we shall demonstrate the existence of a dilatonic-wall
solution, which, under an appropriate bulk
coordinate transformation, results in a
linear dilaton background $\sigma$-model conformal field theory.
Such a solution exhibits holographic properties, which may be used
to provide a resolution of the bulk naked singularity problem.
We shall analyze first the important r\^ole of the higher-curvature
corrections in single brane scenaria, and discuss their physical
disadvantages. Then we shall proceed to analyze the
multibrane scenaria, and discuss their advantages over single brane models.

\section{Four-Dimensional Effective Action}

Observers on brane worlds will have to integrate over the coordinate $z$
in order to obtain the effective four-dimensional action.
The integrated coefficients of the $R^{(4)}(x)$ terms yield
contributions to the effective four-dimensional Planck Mass scale, $M_P$,
whilst the rest of the terms
contribute to the effective four-dimensional vacuum energy.

Using the warped five-dimensional
metric in the form
$ds^2=e^{-2\sigma(z)}\,g^{(4)}_{\mu\nu}(x)dx^\mu\,dx^\nu+dz^2$,
where $\mu,\nu$ are four-dimensional space-time indices
and $g_{\mu\nu}^{(4)}$ is a four-dimensional (brane) metric tensor,
one obtains:
\ba\label{r1}
\sqrt{-g}\,R(x)=
\sqrt{-g^{(4)}(x)}\,\left(e^{-2\sigma(z)}\,R^{(4)}(x)
+e^{-4\sigma(z)}\,{\cal R}\right)
\ea
\ba
\lambda e^{-\frac{4}{3}\Phi(z)} \sqrt{-g}\,R_{GB}(x)&=&
\sqrt{-g^{(4)}(x)}\,\lambda e^{-\frac{4}{3}\Phi (z)}\,
\left(4\,e^{-2\sigma(z)}(3\sigma '(z)^2-2\sigma ''(z))\,
R^{(4)}(x)\right.\nonumber\\
&~&\left.-2\,(R_{\mu\nu\rho\sigma}^2-R_{\mu\nu}^2)+e^{-4\sigma(z)}\,{\cal R}_{GB}\right)
\label{r2}
\ea
\ba
\sqrt{-g}\,(\nabla_\mu\Phi)^2=\sqrt{-g^{(4)}(x)}\,\left(
e^{-4\sigma(z)}\,\Phi'(z)^2+e^{-2\,\sigma(z)}\,(\nabla_i\Phi^{(4)}(x))^2\right)
\label{pp1}
\ea
where
\ba
{\cal R}&=&4\,(5\,\sigma'(z)^2-2\,\sigma''(z))\\
{\cal R}_{GB}&=&24\,\left( 5\,{{\sigma '(z)}^4} -
    4\,{{\sigma '(z)}^2}\,\sigma ''(z)
     \right)
\ea
and the superscript $(4)$ denotes four-dimensional quantities, evaluated on the
brane worlds.

The expression for the four-dimensional
Planck's constant $M_P$, as perceived by an observer living on the
brane world located a finite $z$,
is given by (c.f. (\ref{r1}),(\ref{r2})):
\ba\label{planckexpr}
&~& M_P^2 = M_s^3 \int _{-\infty}^{\infty}
dz  e^{-2\sigma (z)} \left(1 - 4~\lambda
e^{-\frac{4}{3}\Phi (z) }(3(\sigma '(z))^2 - 2 \sigma ''(z))\right)
\ea
where $M_s$ is the five-dimensional (bulk) string mass scale.

In this framework, the
four-dimensional effective vacuum energy
on the observable brane world $\Lambda_{\rm total}$
receives two kinds of contributions:
(i) from the tension of
the brane world we are living on, located, say, at $z=z_i$,
$V_{\rm brane}(z_i) \equiv e^{\omega \Phi (z_i)}v(z_i)$,
and (ii)
from the bulk terms in the action (\ref{s5s4}),
that include the cosmological constant
$\xi$, the dilaton derivative terms,
as well as the ${\cal R}$, ${\cal R}_{GB}$
dependent terms in (\ref{r1}),(\ref{r2}).
Therefore, the expression for the
total four-dimensional vacuum energy
reads:
\ba\label{4dve}
 &~& \Lambda_{\rm total} (z_i)  = \Omega + V_{\rm brane}(z_i)~, \nonumber \\
&~& \Omega = \int_{-\infty}^{+\infty} dz\, e^{-4\sigma (z)}
\left[{\vrule height 13pt width 0pt}\xi e^{\frac{4}{3}\Phi (z)}
-\frac{4}{3} (\Phi '(z))^2 - 20(\sigma '(z))^2  + 8 \sigma
''(z)\right.
+ \nonumber \\
&~&\left.
\lambda e^{-\frac{4}{3}\Phi(z)}\left(24~(5(\sigma '(z))^4 - 4~(\sigma '(z))^2\sigma ''(z)) +
c_2 (\Phi '(z))^4 \right){\vrule height 13pt width 0pt}\right]
\ea

In physically acceptable situations, the quantities
$M_P$ and $\Omega_{\rm total}$
should be finite,
which, in case one encounters
bulk singularities, implies certain
integrability conditions, as we shall discuss later.
This is an important
restriction on model building.

A final remark we would like to make, which we shall make use of in the
following, concerns the induced mass hierarchy on matter
localized on the brane world. As in \cite{RS}, let us concentrate
for simplicity on scalar fields $\phi$ on the brane located at $z=z_i$,
which we assume described the observable world.
By appropriately normalizing the scalars
$\phi \to e^{\sigma (z)}\varphi$,
so that
they have canonical kinetic terms~\footnote{Conformal dilaton factors
$e^{\omega \Phi (z_i)}$ are also present and understood to
be absorbed in the definition of $\varphi$ fields.},
we obtain the following
four-dimensional matter effective action:
\begin{equation}
{\cal S}_m^{(4)} = \int d^4 x \sqrt{-g_{(4)}} \left[
(\partial_\mu \varphi )(\partial_\nu \varphi )g^{\mu\nu}_{(4)} +
m^2 e^{-2\sigma(z_i)} \varphi^2 + \dots \right]
\end{equation}
from which the effective four-dimensional masses are:
\begin{equation}
      m_{\rm obs} = me^{-\sigma(z_i)}
\label{effmasses}
\end{equation}
This accounts for the observed hierarchy in specific cases, as we shall
discuss below~\cite{RS}. For this purpose the mass (\ref{effmasses})
should be compared with the four-dimensional Planck mass (\ref{planckexpr}).
In what follows we shall examine the hierarchy issue in various cases, in
conjunction with stability criteria of the brane worlds,
as well as the problem of the vacuum energy on the observable brane.

\section{Some Exact Solutions to the Equations of Motion}

It was shown in \cite{mr} that the
system (\ref{gravitoneq}),(\ref{dilatoneq}) admits  two
{\it exact}  bulk solutions. Both are characterized by one parameter
($\lambda$),
the string coupling. The first is the original RS solution
with a constant dilaton:
\ba\label{RSexact}
\sigma(z)=\tilde{\sigma}_0+k\,z\ ,\ \Phi={\tilde \phi}_0
\ea
where $\tilde{\sigma_0}={\rm constant}$ and
\ba
k^2=\frac{1}{12\,\lambda}\,e^{4{\tilde \phi}_0/3},  \qquad \xi  =\frac{5}{6\,\lambda}.
\ea
It is important to mention that the RS solution exists {\it only}
for a fixed sign of the higher-curvature parameter $\lambda >0$
(in our
convention) with respect to the Einstein term
in (\ref{s5s4})).  This {\it is the correct sign} implied
by the string-(tree)-amplitude matching procedure
in the bulk~\cite{string}. The solution is indicated by the dark dots
in figure \ref{fig:solutions}.

In this category of linear RS type solutions with constant dilaton
$\Phi=\phi_0 = 0 $ (for brevity)
there is a second type of solution~\cite{mr}
which exists only for $\lambda > 0$:
\ba\label{typeiib}
&~& \zeta=\theta=\omega=0 \nonumber \\
&~& k_- = -\sqrt{\frac{1}{2\lambda} - k_+^2}~,
\qquad \xi = -12k_+^2 (-1 + 2\lambda k_+^2) > 0 ~\qquad({\rm anti~de~Sitter})
\ea
Such a solution is appropriate, in principle, for
effective actions related to three branes embedded
in a bulk higher-dimensional space time, which are known to be solutions
of type II string theories~\cite{bachas}. Indeed, in such cases
the effective low-energy action reads (in the Einstein frame
in the normalization for the dilaton $\varphi$ of ref. \cite{bachas}):
\ba
&~& {\cal S}_{{\rm IIA,B}} = -\frac{1}{2\kappa_{(10)}^2}\int d^{10}x
\sqrt{-g}\left[R + \frac{1}{2}(\partial \varphi )^2 + \frac{1}{12}e^{-\varphi}
(dB)^2 + \sum \frac{1}{2(p+2)!}e^{(3-p)\varphi/2}(dC^{(p+1)})^2 \right]~,
\nonumber \\
&~& {\cal S}_{D_p} = \int d^{p+1}y \left(T^{(p)} e^{(p-3)\varphi/4}\sqrt{-{\hat g}^{(p+1)}}
+ \rho^{(p)}{\hat C}^{(p+1)}\right)
\ea
where $\kappa^2_{(10)}$ is the ten-dimensional gravitational coupling,
$T^{(p)}$ is the $Dp$-brane tension, and the hatted quantities
denote quantities on the brane. The $C$ fields are the Ramond-Ramond fields,
and $\rho^{(p)}$ denotes the corresponding charge density.
We then observe that for three branes $p=3$ the appropriate dilaton
exponential factors become unity, which is compatible with the
solution (\ref{typeiib}) (when translated to the appropriate five-dimensional
case upon dimensional reduction, as discussed in \cite{mr}).
The type II string theory, however, does not have Gauss-Bonnet
interactions at tree $\sigma$-model level, but such terms can be
generated through appropriate string loop corrections. The sign
of such terms though is not definite as yet.
In this paper we shall not be dealing further with the
solution (\ref{typeiib}).

The second exact solution, hereafter called {\it logarithmic},
which we shall be dealing with here,
is
a dilatonic domain wall (bulk naked singularity) of the form:
\ba\label{lowzeesol2}
\sigma(z)=\sigma_0+\sigma_1\, \log|1-\frac{z}{z_s}| \ , \
\Phi(z)=\phi_0-\frac{3}{2} \log|1-\frac{z}{z_s}|
\ea
where
\ba
&~& \hat\lambda\equiv \frac{e^{-4\phi_0/3}\lambda}{z_s^2} =\frac{8(1 +{{\sigma }_1}) }
  { 27 c_2 - 64\,{{{\sigma }_1}}^2 +
      96\,{{{\sigma }_1}}^3  } \nonumber \\
&~& \hat\xi\equiv z_s^2\,e^{4\phi_0/3}\,\xi  =
\frac{3\,\left( 27\,{c_2} +
      81\,{c_2}\,{{\sigma }_1} +
      8\,\left( 16 + 27\,{c_2} \right)
         \,{{{\sigma }_1}}^2 +
      832\,{{{\sigma }_1}}^3 +
      384\,{{{\sigma }_1}}^4 +
      640\,{{{\sigma }_1}}^5 \right) }
    {54\,{c_2} -
    128\,{{{\sigma }_1}}^2 +
    192\,{{{\sigma }_1}}^3} \nonumber \\
&~&
\label{lowzeesol3}
\ea
and $\sigma_0, \sigma_1$ are constants.
This solution is indicated by the dashed line in
figure \ref{fig:solutions}.

For this solution
there are potentially dangerous contributions to
the four-dimensional Planck Mass $M_P$ (\ref{planckexpr}) and
vacuum energy on the observable world $\Omega_{\rm total}$ (\ref{4dve}).
The potentially dangerous contributions
to the vacuum energy
come from the region of $z$ integration
near the location of the naked singularity
$z \sim z_s $~\footnote{It goes without saying that
in the case of naked singularities the $z$-integration
in effective four-dimensional quantities, such as $\Omega$ {\it etc.},
will
not extend to infinity, but bounded by the location
$z_s$ of the singularities, since the latter operate as
completely impenetrable walls, and thus cut off the bulk space.}.
In order to ensure the finiteness
of the brane cosmological constant
we constrain $\sigma_1$ such that
\ba
\sigma_1 < -1/4.
\label{finiteness}
\ea
which, as can be readily seen, guarantees
the finiteness of the four-dimensional Planck
mass (\ref{planckexpr}).

We solve numerically
$\sigma_1$ in (\ref{lowzeesol3})
in terms of $\hat\lambda$.
For simplicity we restrict our discussion to the
case $c_2 =0$. The general case does not affect qualitatively
our discussion.
The plot is shown in Figure \ref{fig:lambda}.
First we note demanding positivity of $\lambda > 0$,
leads automatically to
\ba\label{cond}
\sigma_1<-1\, \qquad (c_2=0)
\ea
which guarantees the finiteness of $M_P$, and $\Omega_{\rm total}$,
as we have seen above.

Interestingly enough
we observe, following the discussion below (\ref{lowzeesol2}),
that these constraints imply an {\it anti-de-Sitter type bulk},
i.e. $\xi <0$.

At this point we should mention that,
as far as we are aware of, in the literature~\cite{naked,naked2}
there has been a discussion on logarithmic solutions of the
form (\ref{lowzeesol2}), but
with $\sigma_1=1/4$ which does not guarantee the effective integrability
of the associated singularity, unlike our case here, where such a possibility
is guaranteed by the presence of the higher-curvature terms in the action.

As seen
figure \ref{fig:lambda}, for the entire
physically acceptable range  $\sigma_1<-1, \hat\lambda>0$
we have that $\hat\lambda<<1$.  Thus,
the truncation of our higher-curvature terms in the action
to ${\cal O}(\alpha ')$, to which we
restrict ourselves~\cite{mr}, is self consistent.
In this case one can perform an expansion
in powers of ${\hat \lambda}$
and get some
analytic formul\ae \,:

\begin{figure}[htb]
\begin{center}
\epsfig{figure=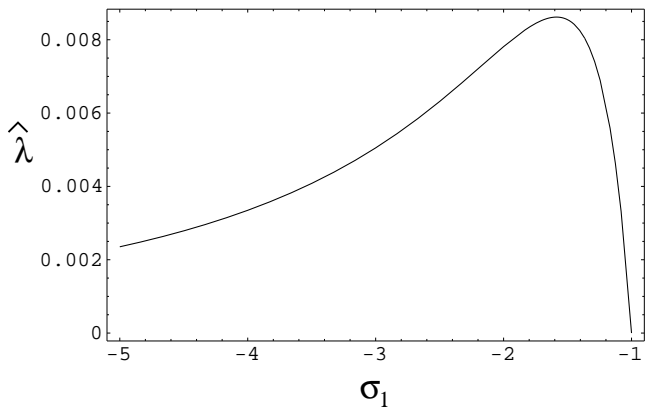, width=0.5\textwidth}
\caption{{\it The dimensionless Gauss-Bonnet coupling
${\hat \lambda}$ versus $\sigma_1$ for $c_2 =0$
(in the allowed regime imposed
by the requirement of integrability of the naked singularities).
We observe that a maximum value of ${\hat \lambda} \sim 8 \times 10^{-3}$
is attained, whilst ${\hat \lambda} \rightarrow 0^+$ as $\sigma _1 \rightarrow
-\infty$.}}
\label{fig:lambda}
\end{center}
\end{figure}

\ba
\sigma_1=-\frac{1}{2\,{\sqrt{3}}\,
     {\sqrt{{{\hat\lambda }}}}} +
  \frac{5}{6} +
  {\cal O}\left({{\hat{\lambda }}}^
   {\frac{1}{2}}\right)\label{ex1}
\ea
or
\ba
\sigma_1=
-1 + \left( -20 + \frac{27\,c_2}
      {8} \right) \,{{\hat\lambda }}
 +
  {\cal O}\left({{\hat{\lambda }}}^
   {2}\right)\label{ex2}
\ea
corresponding to the two branches of Figure \ref{fig:lambda}.
Similarly, we can solve for the bulk cosmological constant
\ba
e^{4\phi_0/3}\,\xi=\frac{5}{6\,{\hat{\lambda }}} -
  \frac{44}
   {3\,{\sqrt{3}}\,{\sqrt{{\hat\lambda }}}}+
 {\cal O}\left({{\hat{\lambda }}}^
   {0}\right)
\ea
or
\ba
e^{4\phi_0/3}\,\xi=9 + \left( 250 - \frac{675\,c_2}
      {16} \right) \,{\hat{\lambda }}+
 {\cal O}\left({{\hat{\lambda }}}^
   {2}\right)
\ea
We also remark that the results presented here, and in later sections,
do not depend crucially on the value of the coefficient
$c_2$ of the quartic-derivative dilaton terms in the action.
Qualitatively the behaviour is virtually independent
of any specific values of $c_2$.
It is for this reason that we keep this parameter generic
in our discussion below,
although we should bear in mind that
the physical situation corresponds to the case where $c_2$
takes on the value (\ref{lambdastring}) compatible with bulk string-amplitude
computations~\cite{string,mr}.

The solution (\ref{lowzeesol2})
leads to a naked singularity in the bulk located at $z=z_s$.
We stress once more that, as we have seen above, by demanding positivity of $\lambda$, as dictated by
tree-level
string amplitude computations~\cite{string}
\footnote{We stress, however, that string loop corrections do not
lead to a definite sign for the Gauss-Bonnet combination. Nevertheless,
for weakly coupled strings, we are dealing with here, such corrections
are subleading, except for the case of type IIB string,
where the tree-level Gauss-Bonnet combination is absent.},
we always obtain {\it Anti-de-Sitter} bulk space-times.
Moreover, from the definition of ${\hat \lambda}$ we obtain the useful relation
\begin{equation}\label{lambdags}
(z_s M_s)^2 =\frac{1}{8g_s^2 {\hat \lambda}}
\end{equation}
where we remind the reader the maximum possible
value of ${\hat \lambda} < 8 \times 10^{-3} $ (cf. figure \ref{fig:lambda}).
Notice that macroscopic distances $z_s \gg 1/M_s$ are attained
for either very small string couplings or in the
limit ${\hat \lambda} \to 0^+$ (or ,equivalently, $\sigma_1 \to -\infty$).

\section{Matching Exact Solutions and the Mass Hierarchy and Cosmological Constant Problems}

\subsection{The Matching Conditions}

It is the purpose of this article
to construct physically interesting
solutions by matching these two exact solutions in various regions
of a bulk spacetime, separated by brane configurations.
Such Brane Junctions, will divide the bulk space
into regions with different, in general, vacuum energies $\xi_i$.
This is physically acceptable~\cite{csaki}, given that the branes operate as
non-trivial sources, upon which four-(space-time)dimensional
matter is restricted to propagate, thereby contributing to the
four-dimensional vacuum energy. We denote such vacuum energies
on the branes, located at $z=z_i$, by $v(z_i)$.
Such brane vacuum energies
are important when matching the solutions (\ref{RSexact}),(\ref{lowzeesol2})
at the various brane junctions, as we shall see below.

The independent equations that enter the matching procedure
read then (with $\varepsilon \to 0^+$):
\begin{eqnarray}
\left[\vrule height 0.6truecm width 0truecm 3\,\sigma '(z) - 4\,\lambda\,e^{\theta\,\Phi(z)}\,\sigma '(z)^3 +
12\,e^{\theta\,\Phi(z)}\,\theta\,\lambda\,\sigma '(z)^2\,\Phi '(z)\right]_{z_i
-\varepsilon}^{z_i + \varepsilon}&=& -\frac{1}{2}\int_{z_i -
\varepsilon}^{z_i + \varepsilon}e^{\omega\,\Phi(z)}\,v(z) dz,
\label{match1} \\
\left[\vrule height 0.6truecm width 0truecm\frac{8}{3}\,\Phi '(z) - 32\,\theta\,\lambda\,
e^{\theta\,\Phi(z)}\,
\sigma '(z)^3 - 4\,c_2\,\lambda\,e^{\theta\,\Phi(z)}\,\Phi '(z)^3
\right]_{z_i - \varepsilon}^{z_i + \varepsilon}&=&-\omega
\int_{z_i-\epsilon}^{z_i+\varepsilon}e^{\omega\,\Phi(z)}\,v(z)dz.
\label{match2}
\end{eqnarray}
It is a solution to these equations (if it exists) that provides
a consistent matching between exact solutions
in various scenarios  which we shall describe below.

In this section we shall analyze various scenarios involving
the exact Randall/Sundrum (RS) solution (\ref{RSexact}) and/or
{\it integrable} bulk naked singularities coming from the
logarithmic solution (\ref{lowzeesol2}).
We shall match the two exact solutions
at brane worlds,
and discuss the physical significance of such constructions.
We shall examine key issues, such as, the resolution of the hierarchy
problem by assuming that our world is a brane embedded in the
bulk, and the cosmological constant problem from the point of view
of four-dimensional observers confined on the brane world.

An important technical issue is the matching of the various
exact solutions in regions of the bulk spacetime
where brane sources exist.
The matching procedure involves integrating the appropriate
bulk equations of motion about the locations $z_i$ of each brane,
and then determining the vacuum energy $v(z_i)$
on each brane in terms of the bulk vacuum energies $\xi_i$
in the various regions.
The relevant matching equations are given in (\ref{match1}),(\ref{match2}).

We distinguish various cases/scenaria, and describe
their advantages or disadvantages, until we arrive at a
physically desirable, and more or less  realistic
multibrane scenario,
where the cosmological constant problem
is resolved, which will be discussed at the end
of our article.

\subsection{A Single Brane RS scenario}

Consider first the case of constant dilaton
and linear $\sigma(z)$ $(\sigma ''(z)=0$).
Setting $\Phi=0$ for simplicity,
the equations of motion in the bulk take the form
(in a generic, but self-explanatory, compact notation):
\ba
(g_{\mu\nu})&:&-3\,R +\lambda\,R_{GB}+5\,\xi=0\\
(\Phi)&:&\lambda\,R_{GB} - \,\xi=0
\ea
The bulk Lagrangian is
\ba
{\cal L}_b=-R+\lambda\,R_{GB}+\xi
\ea
It can then be easily seen that the equations of motion demand
\ba
\frac{1}{2}\,R=\lambda\,R_{GB}=\xi
\ea
and thus the on-shell bulk Lagrangian vanishes ${\cal L}_b=0$.
This is an interesting formal bulk property of the RS solution
in the higher-curvature corrected case, which comes from the
dilaton ($\Phi $) equation of motion.
This should be contrasted with the situation in the standard
RS scenario~\cite{RS}, where ${\cal L}_b =-2\xi/3$.

\begin{figure}
\begin{center}
\epsfig{figure=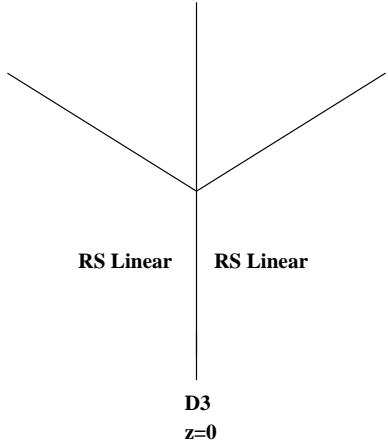, width=0.3\linewidth}
\caption{{\it A single brane (with positive tension) separating the bulk spacetime in two regions
where the exact Randall-Sundrum (linear) solution (\ref{RSexact})
is valid. }}
\label{fig:linear}
\end{center}
\end{figure}

Let us now consider a situation in which one matches
an exact RS solution (\ref{RSexact}), in the two regions of bulk spacetime
separated by a {\it single} brane world at $z=z_0$, in the presence
of higher-curvature string gravity. The scenario is depicted in fig.
\ref{fig:linear} and it involves a warp factor with positive
positive RS parameter $k>0$.

At this point we remind the reader that, in case only the Einstein
term in the five-dimensional action was taken into account~\cite{RS},
such scenaria could be used for a resolution of the four-dimensional
cosmological constant problem, in the sense that the solution
to the Einstein's equations imposed a cancellation of the contributions
of the bulk cosmological constant as seen from the point of view of a
four-dimensional observer and the vacuum energy on the brane.
This solution of course was not better than a fine tuning of the
brane tension.

However, it is straightforward to see that in our case, when
higher-curvature Gauss-Bonnet contributions are taken into account
in the five-dimensional effective action, even such a fine tuning
possibility breaks down, and thus the cosmological constant problem
on the brane remains unsolved in the single brane scenario of
fig. \ref{fig:linear}.

To this end we notice that
the effective four-dimensional Planck constant and bulk cosmological constant
will be given
by (\ref{planckexpr}) and (\ref{4dve})), respectively.
The integral can be split into two parts, $(-\infty, z_0)$ and
$(z_0, +\infty)$, with an appropriate matching
of the solutions at $z=z_0$. It is important to notice that
the scenario of figure \ref{fig:linear}
implies a discontinuity of $\sigma'$ at $z=z_0$, which makes
$\sigma''(z)$ ill defined at $z=z_0$. Without loss of generality,
from now on we take
$z_0=0$.
The most appropriate treatment is to integrate by parts
terms involving $\sigma''$ in effective $z$-integrated quantities
of four-dimensional observers.
Thus,
the
effective four-dimensional
Planck mass (\ref{planckexpr}),
as measured by an observer on the brane at $z=0$,
is:
\ba
M_P^2 =\frac{4}{3}e^{-2{\tilde \sigma}_0}\frac{M_s^3}{k} > 0, \qquad {\rm for} \quad k >0.
\ea
On the other hand, it can be readily seen that the
non-trivial contributions to the bulk cosmological constant as seen by a
four-dimensional observer for the solution (\ref{RSexact})
come only from the discontinuity part involving $\sigma'' (z) (\sigma '(z))^2$.
Treating it with care, as mentioned above, one obtains:
\ba
&~&  \Omega_{{\rm bulk}} =
\int_{-\infty}^{+\infty}~dz~e^{-4{\tilde \sigma}_0-4k|z|}
8\sigma''\left(1-12(\sigma')^2\lambda \right) =
+e^{-4{\sigma}_0}\frac{32}{3}k > 0, \qquad {\rm for} \quad k > 0.
\ea
Above we have used $\sigma' =k$ away from $z=0$.

On the other hand,
the contribution from the brane
tension $-v(z_0)$,
which in this case turns out to be positive~\cite{mr}, is:
\ba
    v(z_0) = -e^{-4{\tilde \sigma}_0}\frac{32}{3}k < 0
\ea
i.e.
the total cosmological constant {\it vanishes}:
\ba
\Lambda_{\rm total} = \Omega_{bulk} + v(z_0(=0)) =0
\ea
Thus, the resolution of the
cosmological constant problem for a
four-dimensional observer which has been provided by a fine tuning
at the level of the Einstein term in the single-brane scenario
of \cite{RS} is still preserved
under the inclusion
of higher-curvature stringy contributions.
However, in this case the value of $k$ is restricted (\ref{RSexact})
to be proportional to the string coupling by means of the
equations of motion. This is a challenge for microscopic
string theory models.
Recall that single brane scenaria
cannot solve the mass hierarchy problem.
On the other hand,  in these single RS scenaria,
the world sits on a positive tension brane, and thus there are
no instabilities.

At this point, however, we stress again that
in our string-inspired models there is a severe restriction,
imposed by the higher-order stringy corrections, namely the
fact that the vacuum energy (or brane tension) turns out (cf (\ref{RSexact}))
to be proportional to the string coupling $g_s$.
This is an exclusive feature of the presence of higher-curvature
stringy corrections~\cite{mr}, and constitutes a challenge for
microscopic string theory brane world models.
At present it is not clear whether such consistent
microscopic models exist.

\subsection{A Single Brane in a Bulk, surrounded by
Integrable Naked Singularities (Dilatonic Domain Walls)}

A second scenario involves the matching of the logarithmic
solution (\ref{lowzeesol2}) in both regions of the bulk spacetime,
separated by a brane world at $z=0$.
In this case there will be naked singularities located at $z=z_s$ and
$z=z_s'$ on both sides of the brane, which will imply a dynamical restriction
of the effective bulk spacetime only within the region between the
singularities (see fig. \ref{fig:linear}). However, we impose the restrictions
(\ref{finiteness}), which guarantee the {\it integrability} of the
naked singularities from the point of view of a four-dimensional
observer.

\begin{figure}
\begin{center}
\epsfig{figure=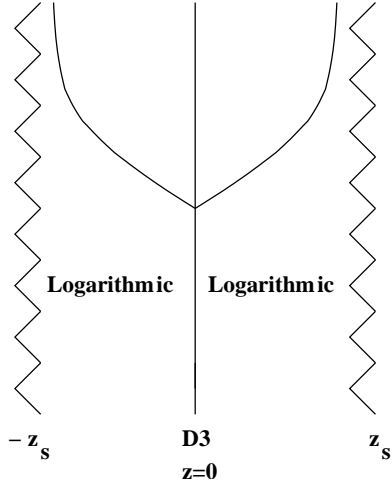, width=0.3\linewidth}
\caption{{\it A single brane (with positive tension) separating the bulk spacetime in two regions
where the exact logarithmic solution (\ref{lowzeesol2})
is valid. }}
\label{fig:log}
\end{center}
\end{figure}

For convenience we shall consider a symmetric scenario, i.e.
a scenario in which the naked singularities are symmetrically positioned
around the brane world at $z=0$.
The complete solution in this case is
\ba
\sigma(z)&=&\,\,\sigma_0+\sigma_1\,\log\left(1-\frac{|z|}{z_s}\right)\\
\phi(z)&=&\phi_0-\frac{3}{2}\,\log\left(1-\frac{|z|}{z_s}\right)
\ea
Solving the matching conditions (\ref{match1}),(\ref{match2})
 and making use of the constraints
(\ref{lowzeesol3}) we find
\ba
\omega&=&\frac{2}{3} \\
z_s\,e^{\omega\,\phi_0}v_0&=&-4\,{{\sigma }_1}\,
  \left( -3 + 4\,{\hat\lambda}\,
     \left( -6 + {{\sigma }_1} \right)
       \,{{\sigma }_1} \right)
\ea
where $v_0$ denotes
the vacuum energy on the brane.
Using the expansions (\ref{ex1}),(\ref{ex2}) we obtain

\ba
e^{\omega\,\phi_0}v_0=\frac{-16}
  {3\,
    {\sqrt{{3\,\hat\lambda}}}}+{\cal O}\left({\hat \lambda}^{0}\right)
\ea
or
\ba
e^{\omega\,\phi_0}v_0=-12+{\cal O}\left({\hat \lambda}^{1}\right)
\ea
We now observe that
the quantity
$v_0$ is negative (thus implying a positive tension brane) in all the allowed
parameter range.
Using (\ref{planckexpr})
\ba
e^{2\sigma_0}\,z_s^{-1}\,\frac{M_p^2}{M_s^3}=\frac{-2\,\left( 1 +
      4\,{\hat\lambda}\,
       \left( -4 + {{\sigma }_1}
         \right) \,{{\sigma }_1}
      \right) }{-1 + 2\,{{\sigma }_1}}
\ea
which is always positive definite. The relevant expansions are
\ba\label{mp1}
e^{2\sigma_0}\,z_s^{-1}\,\frac{M_p^2}{M_s^3}=
  \frac{8\,{\sqrt{{\hat\lambda}}}}
  {{\sqrt{3}}} +
{\cal O}\left({\hat \lambda} \right)
\ea
or
\ba\label{mp2}
e^{2\sigma_0}\,z_s^{-1}\,\frac{M_p^2}{M_s^3}=
\frac{2}{3} +{\cal O}\left(\sqrt{{\hat \lambda}}\right)
\ea
Coming to the total cosmological constant as seen by an observer on the brane,
\ba\label{ltotal}
\Lambda_{\rm total} &=& \Omega (\lambda) + e^{\omega \Phi(0)}\,v_0=\nonumber\\
&~&\frac{4\,\left( -1 +
      e^{4\,{{\sigma }_0}} \right) \,
    {{\sigma }_1}\,
    \left( 81\,{c_2} +
      192\,{{\sigma }_1} -
      32\,{{{\sigma }_1}}^2 +
      256\,{{{\sigma }_1}}^3 \right) }
    {e^{4\,{{\sigma }_0}}\,{z_s}\,
    \left( 27\,{c_2} -
      64\,{{{\sigma }_1}}^2 +
      96\,{{{\sigma }_1}}^3 \right) }
\ea
Normalizing $\sigma_0=0$ we find that the
cosmological constant $\Lambda_{\rm total}$
{\it vanishes} independently
of $\sigma_1$ and $z_s$.

Recall that $z_s $ should be taken to be larger
than $\ell_s=1/M_s$, the string scale, which is the minimum
uncertainty length in string theory.
For the branch  (\ref{mp1}), taking into account (\ref{lambdags}),
we observe that
\begin{equation}
\left( \frac{M_P}{M_s} \right)^2=\frac{8}{g_s \sqrt{3}}
\end{equation}
which implies that $M_s \sim 100$ TeV,
required probably for a superstring resolution of the
Higgs stabilization problem,
are compatible with
$M_P \sim 10^{19}$ GeV for extremely weak string couplings $g_s$.

On the other hand, for the case (\ref{mp2}),
one obtains:
\begin{equation}
\left( \frac{M_P}{M_s} \right)^2 =\frac{2}{3} z_sM_s
\end{equation}
which, on account of (\ref{lambdags}), implies that
TeV string scale is obtained for
$z_s M_s \sim 10^{28}$, which implies
{\it either} weak string coupling
{\it or} small positive values of ${\hat \lambda}$
(i.e. $\sigma_1 \to -\infty $).

This scenario has the advantage
of using a positive tension (observable) brane world,
which is free from instabilities.
However, on account of (\ref{effmasses}), the hierarchy
factor is unity for a brane located at $z=0$ .
Hence
the hierarchy problem between Planck and electroweak scales
is {\it not solved} in this case.
There is, however,
another serious drawback in such constructions in that the
positive tension world faces directly naked singularities.
Although integrable, one could argue, in agreement
with the discussion in \cite{davis}, that
the presence of naked not-shielded singularities
will render the quantum version of such models inconsistent.
For instance, the propagation of gravitational waves in the
vicinity of time-like singularities is not well-defined in the sense
that the time evolution of wave packets is not uniquely defined
in such regions. Moreover,
for the case of null classical singularities~\cite{davis},
there is absorption of incoming radiation, which is also phenomenologically
troublesome.

For these reasons a physically desirable
scenario would be to shield the naked singularities
from the physical brane world
by placing shadow brane worlds in front of them~\cite{naked2}.
Apart from shielding their effects,
this would also allow
for the bulk space between our world and the shielding branes
to accept supergravity  solutions~\cite{davis}.
As we shall show later on,
such solutions yield consistent and physically
acceptable scenaria in the context of our higher-curvature gravity,
allowing simultaneously for a resolution of both the
hierarchy and the cosmological constant problem on our
observable brane world.
To understand, however, such shielding processes
it is instructive to discuss one more single-brane case, in which
the brane faces a single naked bulk singularity. This is done in the
following subsection.

\subsection{Matching Logarithmic and RS (Linear) solutions}

\begin{figure}
\begin{center}
\epsfig{figure=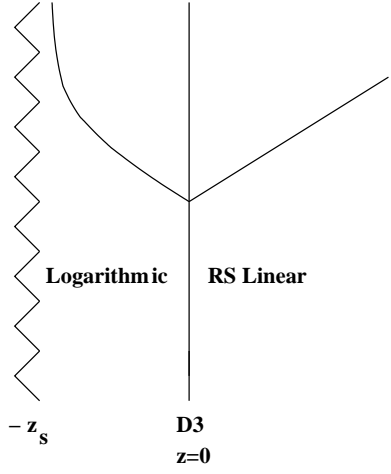, width=0.3\linewidth}
\caption{{\it A single brane (with positive tension) separating the
bulk spacetime in two regions, in one of which
the exact logarithmic solution (\ref{lowzeesol2})
is valid, while in the other the exact linear RS solution
(\ref{RSexact}) holds.}}
\label{fig:loglinear}
\end{center}
\end{figure}

Consider the situation depicted in figure \ref{fig:loglinear}.
The  brane
separates
two regions of the bulk space time,
in which we shall match a logarithmic solution,
with the dilatonic domain wall located at $z=-z_s <0$, with a
linear RS solution.
We discuss first the case, where
the
brane world is placed to the right of the naked singularity, i.e. at
$z=-z_0$ where $z_0>0$ and
$z_0<z_s$ (cf figure \ref{fig:loglinear}). The solution has the form:
\ba
\sigma(z)=\left(\sigma_0+\sigma_1\,\log\left(1+\frac{z}{z_s}\right)\right)\,\vartheta(-z)+
\left(\tilde{\sigma_0}+k\,z\right)\vartheta(z)
\ea
and similarly for $\Phi$. In this section we keep the location of the
brane $z_0$ arbitrary, because the results will be used
later on in the multibrane scenario.

Solving the matching conditions we now obtain
\ba
\omega=\frac{2}{3}, \qquad
(z_s-z_0)\,e^{\omega\phi_0}\,
v=\frac{-8}{3\,{\sqrt{3}}\,{\sqrt{{\hat\lambda}}}} +
  6\,{{\sigma }_1} -
  8\,{\hat\lambda}\,\left( -6 + {{\sigma }_1}
     \right) \,{{{\sigma }_1}}^2
\ea
For the tension we have the expansions
\ba
(z_s-z_0)\,e^{\omega\phi_0}\,v=-\frac{16}{3\,{\sqrt{3}}\,{\sqrt{{\hat\lambda}}}}
+ {\cal O}\left(\hat\lambda^{0}\right)
\ea
or
\ba
(z_s-z_0)\,e^{\omega\phi_0}\,v=-\frac{8}{3\,{\sqrt{3}}\,{\sqrt{{\hat\lambda}}}}
+ {\cal O}\left(\hat\lambda^{0}\right)
\ea
The four-dimensional Planck scale is finite and positive
\ba
e^{2\sigma_0}\,\frac{M_p^2}{M_s^3}&=&\frac{\left( {z_0} - {z_s} \right) \,
    \left( 3 + 4\,{\sqrt{3}}\,
       {\sqrt{{\hat\lambda}}}\,
       \left( 1 - 2\,{{\sigma }_1} \right)  +
      12\,{\hat\lambda}\,
       \left( -4 + {{\sigma }_1} \right) \,
       {{\sigma }_1} \right) }{3\,
    {\left( 1 - \frac{{z_0}}{{z_s}} \right) }^
     {2\,{{\sigma }_1}}\,
    \left( -1 + 2\,{{\sigma }_1} \right) }
\ea
The perturbative expansions in powers of ${\hat \lambda} \ll 1$
are given by:
\ba\label{mp1log}
e^{2\sigma_0}\,z_s^{-1}\,\frac{M_p^2}{M_s^3}=
  \frac{8\,{\sqrt{{\hat\lambda}}}}
  {{\sqrt{3}}} +
{\cal O}\left({\hat \lambda} \right)
\ea
or
\ba\label{mp2log}
e^{2\sigma_0}\,z_s^{-1}\,\frac{M_p^2}{M_s^3}=
\frac{1}{3} +{\cal O}\left(\sqrt{{\hat \lambda}}\right)
\ea

The four-dimensional cosmological constant
(taking into account bulk and brane contributions)
acquires the form:
\ba
\Lambda_{\rm total} &=&
\frac{-8\,{{z_s}}^{4\,{{\sigma }_1}}\,
     {\left( -{z_0} + {z_s} \right) }^
      {-1 - 4\,{{\sigma }_1}}\,
     \left( {\sqrt{3}} -
       6\,{\sqrt{{\hat\lambda}}}\,{{\sigma }_1}
       \right) \,\left( -1 +
       {\sqrt{3}}\,{\sqrt{{\hat\lambda}}}\,
        {{\sigma }_1} +
       6\,{\hat\lambda}\,{{{\sigma }_1}}^2 \right)
       }{9\,{\sqrt{{\hat\lambda}}}\,
     e^{4\,{{\sigma }_0}}} +\nonumber \\
&~&
  \frac{\frac{8\,{\sqrt{3}}}
      {{\sqrt{{\hat\lambda}}}} -
     54\,{{\sigma }_1} +
     72\,{\hat\lambda}\,
      \left( -6 + {{\sigma }_1} \right) \,
      {{{\sigma }_1}}^2}{9\,
     \left( {z_0} - {z_s} \right) } \nonumber  \\
&~&-
  \frac{{{z_s}}^{4\,{{\sigma }_1}}\,
     {\left( -{z_0} + {z_s} \right) }^
      {-1 - 4\,{{\sigma }_1}}\,
     \left( 16\,\left( -3 + {\hat\xi} -
          8\,{{\sigma }_1} - 20\,{{{\sigma }_1}}^2
          \right) +
       3\,{\hat\lambda}\,
        \left( 27\,{c_2} +
          512\,{{{\sigma }_1}}^3 +
          640\,{{{\sigma }_1}}^4 \right)  \right)
     }{16\,e^{4\,{{\sigma }_0}}\,
     \left( 1 + 4\,{{\sigma }_1} \right) } \nonumber \\
\label{loglincosmo}
\ea
One observes that $\Lambda_{\rm total}$
vanishes for
\ba
\sigma_0=-\sigma_1\,\log\left(1-\frac{z_0}{z_s}\right)
\label{condition}
\ea

We next remark on the case in which
the dilatonic wall (naked singularity) lies to the right of the
brane world, i.e
the brane lies at $z=z_0>0$ and the singularity at $z=z_s>0$,
such that $z_s-z_0 >0$.
The solution for the metric function assumes the form:
\ba
\sigma(z)=
\left(\tilde{\sigma_0}-k\,z\right)\vartheta(-z)+
\left(\sigma_0+\sigma_1\,\log\left(1-\frac{z}{z_s}\right)\right)\,\vartheta(z)
\ea
A similar analysis as above shows that
the resulting brane tension, the quantity $\omega$, and the cosmological
constant turn out to be exactly the same
as in the previous case of figure \ref{fig:loglinear}.

\subsection{Multibrane Scenaria with RS solutions and
Brane-Shielded
Naked Singularities}

\begin{figure}
\begin{center}
\epsfig{figure=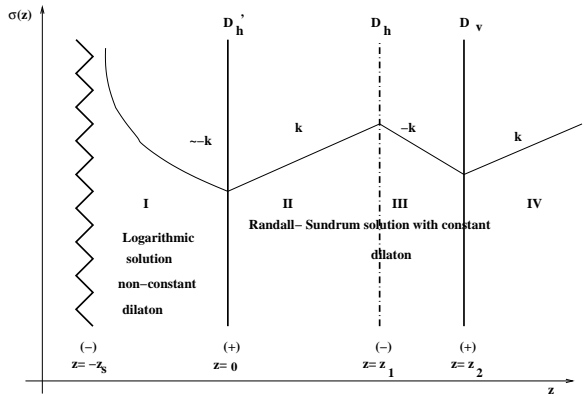, width=0.45\linewidth}
\caption{{\it
A multibrane scenario, in which our world is represented
as a positive tension brane (at $z=z_2$), having from the left branes with
alternating-sign tensions, which shield a
bulk naked singularity (which may be thought of
as a limiting (singular) case of a negative tension brane).
To the right of the brane world,
on the other hand,
the bulk dimension extends to infinity.}}
\label{fig:multibrane1}
\end{center}
\end{figure}

We are now well equipped to discuss  multibrane
scenaria, in which our world is a positive tension brane,
and
the naked singularities are shielded
by ghost brane worlds.
The first of these scenaria
is depicted
in figure \ref{fig:multibrane1}.
The scenario is asymmetric, and to the left of our world
there exist alternating tension branes that shield
a bulk naked singularity, located at $z=-z_s$.

The reader should notice that
in calculating the Planck Mass
$M_P$ and the total Vacuum energy on the brane world $\Lambda_{\rm total}$
one can use directly the results presented in the previous
sections.
In regions I and II of figure \ref{fig:multibrane1} one considers a
matching between a logarithmic and a linear RS solution,
in
regions II and III one matches two linear RS solutions
at a negative-tension brane, while in regions II and IV
one matches two RS solutions at a positive-tension brane (our world).

As far as the Planck mass (\ref{planckexpr})
is concerned, the non-trivial
contributions come from the logarithmic-linear region (I and II),
as discussed in the relevant subsection previously.
The remaining of $z$-integration over the regions II,III and IV
yield cancelling contributions to $M_P$.
The final result of $M_P$ therefore is given by (\ref{mp1log}),
(\ref{mp2log}), for the two cases.

The cosmological constant on the other hand, as discussed previously,
receives contributions only from the
region I and from the
brane discontinuities at $z=0, z_1, z_2$
(cf. figure \ref{fig:multibrane1}).
The contributions from $z_1$ and $z_2$ cancel, thereby leaving one
with the result of the previous subsection, 
which allows for {\it vanishing}
total cosmological constant on the brane world, by appropriately
choosing the constant $\sigma_0$ of the metric function $\sigma(z)$.

Moreover, the hierarchy factor on our world for this scenario
is given by $e^{-k(2z_1-z_2)}$, thus allowing for a resolution of the
hierarchy even on a positive tension brane, as in the scenario of
\cite{bigravity}. To obtain the desired hierarchy one needs
\begin{equation}
 k(2z_1 - z_2) ={\cal O}(50)~, \qquad k=\sqrt{\frac{2}{3}}g_s M_s
\end{equation}
where we took into account the property of the exact RS solution
(\ref{RSexact})
which connects $k$ to the string coupling $g_s < 1$~\cite{mr}.

Notice that the hierarchy is solved only in the asymmetric case
i.e. when $z_2 > 2z_1$. In the orbifold case there is no natural
explanation why such an asymmetry should occur in nature.
In our case,
on the other hand,
where a matching of two different
exact solutions occurs, the asymmetry between $z_1, z_2$
is quite natural in our opinion. Indeed, the presence of
a bulk naked singularity guarantees the existence of
a strong gravitational attraction between the shielding branes
and the bulk naked singularity.
One therefore obtains the desired hierarchy in the case
where the shielding branes are close (relative to the string scale
$1/M_s$) to the naked singularity, while the observable world
lies far away from it.

\begin{figure}
\begin{center}
\epsfig{figure=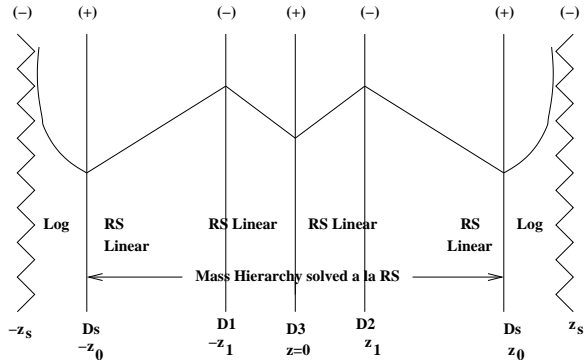, width=0.45\linewidth}
\caption{{\it
A multibrane scenario, in which our world is represented
as a positive tension brane (at $z=0$), surrounded by branes
with alternating-sign tensions
that shield two symmetrically-positioned
bulk naked singularities.}}
\label{fig:multibrane}
\end{center}
\end{figure}

A second multibrane scenario
is depicted
in figure \ref{fig:multibrane}, in which
our world, represented by a positive tension
3-brane, lies between four branes and two naked singularities,
positioned {\it symmetrically}.
The branes are assumed to have alternating-sign tensions, starting
from positive tension branes facing directly
the naked singularity.
This scenario replaces the orbifold scenaria, but in a dynamical way,
given that the presence of naked singularities imply
a restriction of the bulk space time.
The results are quite similar to the previous asymmetric multibrane
case.

In the next section we discuss another interesting scenario
in which the bulk naked singularities are resolved by means of
holography and bulk/boundary correspondence for a specific
dilatonic-domain wall solution.

\section{A Holographic Resolution of the Naked Singularity Problem}

\subsection{Linear Dilaton Exact Solution}

Let us commence our analysis in this
section by considering the Einstein-frame  metric  (\ref{RSmetric})
with the following metric function and dilaton
configurations
\ba
\sigma (z) = \sigma_0 + \sigma_1 {\rm log}|1-\frac{z}{z_s}|, \nonumber \\
\Phi (z) = \phi_0 - \frac{3}{2} {\rm log}|1-\frac{z}{z_s}|
\ea
where $\gamma$ to be determined below.

In this section we consider the special case of the exact solutions
(\ref{lowzeesol2}), (\ref{lowzeesol3})
corresponding to $\sigma_1 =-1$,
which still corresponds to integrability of the naked singularities.
We shall also set $\sigma_0=0$
without loss of generality for brevity.

In this case one finds that this is an exact solution provided that:
\begin{equation}\label{parsol}
c_2 = 160/27~, \qquad
{\hat \lambda} = \frac{1}{52} > 0~, \qquad {\hat \xi}=\frac{171}{26} > 0
\qquad ({\rm anti~de~Sitter~bulk})
\end{equation}

In fact, in this case,
by performing the coordinate transformation
\ba
dz \to dy \equiv \frac{dz}{|1-\frac{z}{z_s}|}
\to y=z_s{\rm log}|1-\frac{z}{z_s}|
\ea
one may cast the Einstein metric in the form:
\ba\label{metric2}
    ds^2 = e^{2\frac{y}{z_s}}\left[ \eta_{ij}dx^i dx^j + (dy)^2 \right]
\ea
and the dilaton is linear in the coordinate $y$:
\ba\label{dilaton}
\Phi = \phi_0 -\frac{3}{2}\frac{y}{z_s}
\ea
We set from now on $\phi_0=0$ by absorbing such constants
in the definition of the string coupling $g_s$.

We now connect the metric (\ref{metric2}) in terms of a
$\sigma$-model metric, $G_{\mu\nu}^\sigma$
by following standard methods~\cite{string}:
\ba\label{einstmetric}
   g_{\mu\nu}^{\rm Einst} =e^{-4\Phi/3}G^\sigma _{\mu\nu} =
e^{2\frac{y}{z_s}}G_{\mu\nu}^\sigma
\ea
in our normalizations,
where $\mu.\nu = 0, \dots 4$. We observe
compatibility with (\ref{metric2}), which
implies that there is a $\sigma$-model metric
which is Minkowski flat, and a linear dilaton background~\cite{aben}
(\ref{dilaton})
in the $y$ variable:
\ba\label{smodelmetric}
&~&   G_{\mu\nu}^\sigma = \eta_{\mu\nu} \nonumber \\
&~& \Phi = - \frac{3}{2}\frac{y}{z_s}
\ea
As we have seen, this is a solution of
the conformal invariance conditions to
order ${\cal O}(\alpha ')$, and probably to
all orders in $\alpha'$, in view of the Coulomb-gas
conformal field theory correspondence~\cite{aben}.

Such a solution is that of a non-critical bulk string
with subcritical central charge deficit
$\delta c = -Q^2 = -\frac{1}{2z_s^2} <0$,
with the r\^ole of the Liouville mode played by the spacelike coordinate $y$.
This implies a linear dilaton (up to an irrelevant constant)
of the form $\Phi = -\frac{3}{\sqrt{2}}Qy$ in our normalization.
As a result of this central charge deficit
there is a dilaton potential (in the Einstein frame)~\cite{aben}
\ba
   V(\Phi) \propto Q^2 \sqrt{-g^E} e^{\zeta \Phi} = \frac{1}{2z_s^2}
\frac{1}{|1-\frac{z}{z_s}|^2} >0,
\ea
where the superscript $E$ denotes quantities in Einstein frame,
hence implying anti-de-Sitter bulk geometries. This is nothing other
than the result (\ref{parsol}) 
for the bulk cosmological constant ${\hat \xi}=\frac{171}{26} > 0$
of our solution, thereby providing a nice consistency check
of the representation of this solution as a non-critical string. 
Notice that the positive (in our convention) sign of
$V$ is due to the fact that the signature of the Liouville field
is space like (subcritical strings with central charge deficit
$\delta c <0$). This is the opposite situation from that of \cite{aben}.

In general, for singularities at a finite distance from a brane world,
the bulk potential can be cancelled by a fine tuning of the brane tension.
As a result of the anti-de-Sitter character of the bulk cosmological constant
one needs a positive tension brane ($-v_i$) for such a cancellation to be
operative.
The potential vanishes, and hence the bulk cosmological constant on our
world vanishes as well, if the bulk singularity is placed at an infinite
distance from our world.
This corresponds to a vanishing central charge
deficit, and hence to a critical string. In this solution the
distance of the brane world from the singularity domain wall, then,
measures the subcriticality of the string in some sense.
In the case of an infinite-bulk-distance singularity
one needs a vanishing brane tension in order
to ensure a vanishing total vacuum energy
on the brane. Otherwise, a positive brane tension may result in a
non-trivial positive (de-Sitter type)
four-dimensional cosmological constant.

\subsection{A Holographic Resolution of the Naked Singularity}

The feature of linear dilaton solutions is that,
if the string coupling was weak, then they would be
{\it holographic}~\cite{aharony}, in the sense that the
bulk theory will be dual to a theory without gravity on the
boundary of the Anti-de-Sitter bulk geometry.
The string coupling $g_s =e^{\Phi} \to \infty$
at the singularity $z \to z_s$, but it is {\it weak}
at a brane world placed at infinity $z \to \infty$, where the cosmological
constant vanishes. Indeed in that case $y \to +\infty$ and the string coupling
$g_s = e^\Phi \to 0$.

The holographic property follows directly from the fact~\cite{adshol}
that in our case
the bulk space is of anti-de-Sitter type (\ref{parsol}).
It can be readily seen also from the form of the
Einstein metric~\cite{aharony} (\ref{einstmetric}), which diverges at
the boundary  $y \to \infty$, thereby implying that the distances
between fixed points
on the boundary diverge, and hence signals should propagate
in the bulk before they interact on the boundary, a necessary
condition for holography~\cite{susskhol}. In an equivalent manner,
at the string frame (\ref{smodelmetric}), the string coupling
$g_s \sim e^{\Phi} \to 0$ as $y \to \infty$, and hence the string interactions
vanish on the boundary, and thus one can see
again, from this point of view, that signals must propagate on the bulk
before they can interact.

This holography implies~\cite{aharony}
that the bulk (five-dimensional theory) with gravity is
dual to a field theory without gravity on a brane world at infinity,
$y \to \infty$, defining the boundary of the anti-de-Sitter bulk space time.
The dual theory on the boundary need not be local.
However, we cannot calculate reliably
at the singularity domain wall. A resolution on the singularity
might be provided by an appropriate world-sheet superpotential
as in two-dimensional Liouville theory~\cite{aharony}.
This is an issue to be looked at
more carefully in our models in a future work.

The above holographic situation implies a non-trivial and
non-perturbative resolution of the naked singularity problem
in this case,
which in this way can be mapped into a bulk/boundary problem.
Moreover we have also seen that the boundary theory has a
vanishingly small cosmological constant.

\section{Conclusions and Outlook}

We have discussed in this work some exact solutions
to the graviton and dilaton equations of motion
in brane world scenaria, in the bulk space of which one assumes
a closed-string gravitational multiplet propagating.
We considered ${\cal O}(\alpha ')$ string effective actions
in five dimensions, with graviton and dilaton fields.
We have ignored antisymmetric tensor fields for simplicity.

We have managed to demonstrate that the (original)
linear Randall-Sundrum scenario
with a single brane, or two branes and an orbifold construction,
can simultaneously solve, both the hierarchy and the
smallness of the cosmological constant. The orbifold  scenario
is free from instabilities associated with negative tension
boundary branes, since in that case such instabilities are projected
out of the spectrum. A challenge for microscopic string theory models,
however,
is to construct theories whose brane tensions (as a result
of matter gauge fields) are proportional to the string coupling
in the way dictated in our solution (\ref{RSexact}).

We have also discussed scenaria which
avoid orbifold compactifications, but involve
bulk naked singularities (though integrable).
We have succeeded in constructing a mathematically
consistent model, depicted in
fig. \ref{fig:multibrane}, in which our world
is a positive tension brane, surrounded by
alternating-sign-tension branes that shield the
effects of two dilatonic domain walls
positioned symmetrically on either
side of our world. We have demonstrated a resolution of the
mass hierarchy simultaneously with the vanishing of the
cosmological constant on the brane.
Our scenario, although involving negative tension branes,
however is different from the bigravity models of \cite{bigravity}
in that here we do not impose orbifold compactification.
The bulk dimension is dynamically cut-off by the presence
of the dilatonic walls, whose distance from the shielding branes
can, in certain models, become
very
large and thus be responsible for the smallness of the four-dimensional
cosmological constant and a resolution of the mass hierarchy
in a geometric fashion.

An important issue, which we did not discuss here,
but we certainly plan to come back to
in a future publication, concerns localization of gravity in such
higher-curvature scenaria.
In the recent literature there has been some
discussion of this
important issue, in the context of purely gravitational
Gauss-Bonnet effective actions, with constant
dilaton fields~\cite{deloc}.
The problem of non-constant bulk dilatons,
which are essential in the scenaria studied
in \cite{mr} and here, remains an open issue.
It is this issue that will show whether the
multibrane scenario of fig. \ref{fig:multibrane}
stands a chance of being a phenomenologically
acceptable model.

Another important issue is the dynamical stability
of the above configurations, especially after the inclusion
of quantum corrections.
Unfortunately, in general, the presence of negative tension branes violates
the energy conditions, e.g. the weakest of them, which
could be stated as~\cite{weakest}:
\ba
  \sigma ''(z) \ge 0
\ea
and should hold globally in the bulk.
A negative tension RS brane, for instance, located at $z_i$
violates
locally this condition, given that $\sigma ''(z_i) = -|k|\delta (z-z_i)
< 0$.  One should therefore expect an instability at some scale,
which for bigravity models involving a negative tension brane between
positive tension ones, as in figure \ref{fig:multibrane},
is manifested through the appearance of a ghost radion mode.
The latter is
a scalar fluctuation mode describing bending of
the brane, whose kinetic term has the wrong sign
below some characteristic separation distance between
the negative and positive tension branes~\cite{radion}.
We should mention that, to avoid the problem
of negative-tension branes,
constructions in six dimensional curved bulk spaces
have been considered~\cite{ross}, which admit
only positive tension brane worlds, as a result of
the non-trivial curvature of the six-dimensional bulk.
This was not possible in five dimensions.
Such models appear to be radion-ghost-free, thus avoiding
the instability, in agreement with the global satisfaction
of the positive-energy conditions in the bulk.
From the point of view of string theory, propagating in the bulk,
considered here and in \cite{mr}, one
should in principle have at his/her disposal the full ten-dimensional
spacetime. It would be interesting, therefore,
to see how the results are affected
in case one considers higher-curvature string-inspired corrections
in such models, along the lines discussed
in the five-dimensional context here.
We plan to come to these issues in a future
work.

We have also discussed a holographic scenario for the
resolution of the bulk naked singularities, in which the
singularities lie infinitely far away from our world, the latter
being viewed as a boundary of an anti-de-Sitter bulk, which is known
to exhibit holographic properties.

Finally, we stress that in our model~\cite{mr}, the
brane tensions turn out to be proportional to the string coupling
$g_s$. This is a quite important restriction, and remains as a challenge
for microscopic string theory models to reproduce this effect
explicitly. It is only in that case that the models constructed
here could be considered as complete and free from any sort of fine tuning.
Actually the fact that our model involves necessarily brane configurations
with opposite brane tensions might be thought of as fine tuning.
As argued in \cite{davis}, however, such configurations appear natural
in {\it bulk supersymmetric} scenaria,
which are expected to occur in realistic superstring models.
From this point of view it is the breaking of
supersymmetry that will induce a tension detuning.
The breaking of supersymmetry, however,
is not a classical phenomenon, and,
thus, in principle, one does not
expect it to introduce any inconsistencies
with the above classical construction.
We plan to return to a detailed study of such issues in due course.

\section*{Acknowledgements}

J.R. wishes to thank CPHT, Ecole Polytechnique (France) and
the Physics Department of King's College London for
hospitality during various stages of this work.
The work is partially supported by the European Union
(contract ref. HPRN-CT-2000-00152).

\end{document}